\documentclass[amsmath,amssymb,showpacs]{revtex4}
\usepackage{epsfig}
\usepackage{graphicx}

\begin{document}

\title{The spacetime structure of MOND with Tully-Fisher relation and Lorentz invariance violation}

\author{Xin Li}
\email{lixin@ihep.ac.cn}
\author{Zhe Chang}
\email{changz@ihep.ac.cn}
\affiliation{Institute of High Energy Physics\\
and\\
Theoretical Physics Center for Science Facilities,\\
Chinese Academy of Sciences, 100049 Beijing, China}

\begin{abstract}
It is believed that the modification of Newtonian dynamics (MOND) is possible alternate for dark matter hypothesis. Although Bekenstein's TeVeS supplies a relativistic version of MOND, one may still wish a more concise covariant formulism  of MOND. In this paper, within covariant geometrical framwork, we present another version of MOND. We show the spacetime structure of MOND with properties of Tully-Fisher relation and Lorentz invariance violation.
\end{abstract}
\pacs{95.35.+d,02.40.-k, 04.25.Nx}

\maketitle
\section{Introduction}
In 1932, Physicists \cite{Oort,Zwicky} found from the observations of galaxies and galaxy clusters that the Newton's gravity could not provide enough force to attract the matters of galaxies. The recent astronomical observations show that  the
rotational velocity curves of all spiral galaxies tend to some
constant values\cite{Trimble}. These include the Oort discrepancy in
the disk of the Milky Way\cite{Bahcall}, the velocity dispersions of
dwarf Spheroidal galaxies\cite{Vogt}, and the flat rotation curves
of spiral galaxies\cite{Rubin}. These facts violate sharply the prediction of Newton's gravity. The most widely adopted way to resolve these difficulties is the dark matter hypothesis. It is assumed that all visible stars are
surrounded by massive nonluminous matters.

The dark matter hypothesis has dominated astronomy and cosmology for almost 80 years. However, up to now, no direct observations has been substantially tested. Some models have
been built for alternative of the dark matter hypothesis. Their main
ideas are to assume that the Newtonian gravity or Newton's dynamics
is invalid in galactic scale. The most successful and famous model is MOND \cite{Milgrom}. It assumed that the
Newtonian dynamics does not hold in galactic scale. The particular
form of MOND is given as
 \begin{equation}
 \begin{array}{l}
  m\mu\left(\displaystyle\frac{a}{a_0}\right)\mathbf{a}=\mathbf{F},\\[0.4cm]
 \displaystyle\lim_{x\gg1}\mu(x)=1,~~~\lim_{x\ll1}\mu(x)= x,
 \end{array}
 \end{equation}
where $a_0$ is at the order of $10^{-8}$ cm/s$^2$. At beginning, as
a phenomenological model, MOND explains well the flat rotation
curves with a simple formula and a new parameter. In particular, it
deduce naturally a well-known global scaling relation for spiral
galaxies, the Tully-Fisher relation\cite{TF}. The Tully-Fisher relation is an empirical relation between the
total luminosity of a galaxy and the maximum rotational speed. The relation is of the form $L\propto v^a_{max}$, where $a\approx4$ if luminosity is measured in the near-infrared. Tully and Pierce \cite{Tully} showed that the Tully-Fisher relation appears to be convergence in the near-infrared. McGaugh \cite{McGaugh} investigated the Tully-Fisher relation for a large sample of galaxies, and concluded that the Tully-Fisher relation is fundamental relation between the total baryonic mass and the rotational speed. MOND \cite{Milgrom} predicts that the rotational speed of galaxy has an asymptotic value $\lim_{r\rightarrow\infty}v^4=GMa_0$, which explains the Tully-Fisher relation.

By introducing several scalar, vector and tensor fields, Bekenstein\cite{Bekenstein}
rewrote the MOND into a covariant formulation (TeVeS). He showed that the
MOND satisfies all four classical tests on Einstein's general
relativity in Solar system. Beside Bekenstein's theory, there are other MOND theories(for example, Einstein-aether theory\cite{Zlosnik}). These MOND theories modify gravity with additional scalar/vector/tensor fields. Bekenstein's theory and Einstein-aether theory both admit a preferred reference frame and break local Lorentz invariance. It means that local Lorentz symmetry violation is a feature of MOND.

The local Lorentz symmetry violation implies that the space structure of galaxy is not Minkowskian at large scale, and the relationship between the Tully-Fisher relation and MOND implies that the space structure of galaxy depends on rotational speed. Finsler gravity based on Finsler geometry involves the above features. It is natural to assume that the Finsler gravity is a covariant formulism of MOND. Finsler geometry \cite{Book by Bao} as a nature extension of Riemann geometry involves Riemann geometry as its special case. The length element of Finsler geometry depends not only on the positions but also the velocities. Finsler gravity naturally preserves fundamental principles and results of general relativity. A new geometry (Finsler geometry) involves new spacetime symmetry. The Lorentz invariance violation is intimately linked to Finsler geometry. Kostelecky \cite{Kostelecky} has studied effective field theories with explicit Lorentz invariance violation in Finsler spacetime. In this paper, within the covariant geometric framework, we try to present a covariant formulism of MOND with explicit Tully-Fisher relation and Lorentz invariance violation.

\section{Vacuum field equation in Finsler spacetime}
Instead of defining an inner product structure over the tangent bundle in Riemann geometry, Finsler geometry is based on
the so called Finsler structure $F$ with the property
$F(x,\lambda y)=\lambda F(x,y)$ for all $\lambda>0$, where $x$ represents position
and $y\equiv\frac{dx}{d\tau}$ represents velocity. The Finsler metric is given as \cite{Book
by Bao}
 \begin{equation}
 g_{\mu\nu}\equiv\frac{\partial}{\partial
y^\mu}\frac{\partial}{\partial y^\nu}\left(\frac{1}{2}F^2\right).
\end{equation}
Finsler geometry has its genesis in integrals of the form
\begin{equation}
\label{integral length}
\int^r_sF(x^1,\cdots,x^n;\frac{dx^1}{d\tau},\cdots,\frac{dx^n}{d\tau})d\tau~.
\end{equation}
The Finsler structure represents the length element of Finsler space.

The parallel transport
has been studied in the framework of Cartan
connection \cite{Matsumoto,Antonelli,Szabo}. The notation of parallel
transport in Finsler manifold means that the length
$F\left(\frac{dx}{d\tau}\right)$ is constant.
The geodesic equation for Finsler manifold is given as \cite{Book by Bao}
\begin{equation}
\label{geodesic}
\frac{d^2x^\mu}{d\tau^2}+2G^\mu=0,
\end{equation}
where
\begin{equation}
\label{geodesic spray}
G^\mu=\frac{1}{4}g^{\mu\nu}\left(\frac{\partial^2 F^2}{\partial x^\lambda \partial y^\nu}y^\lambda-\frac{\partial F^2}{\partial x^\nu}\right)
\end{equation} is called geodesic spray coefficient.
Obviously, if $F$ is Riemannian metric, then
\begin{equation}
G^\mu=\frac{1}{2}\tilde{\gamma}^\mu_{\nu\lambda}y^\nu y^\lambda,
\end{equation}
where $\tilde{\gamma}^\mu_{\nu\lambda}$ is the Riemannian Christoffel symbol.
Since the geodesic equation (\ref{geodesic}) is directly
derived from the integral length
\begin{equation} L=\int
F\left(\frac{dx}{d\tau}\right)d\tau,
\end{equation} the inner product
$\left(\sqrt{g_{\mu\nu}\frac{dx^\mu}{d\tau}\frac{dx^\nu}{d\tau}}=F\left(\frac{dx}{d\tau}\right)\right)$
of two parallel transported vectors is preserved.

In Finsler manifold, there exists a linear connection~-~the
Chern connection \cite{Chern}. It is torsion freeness and almost
metric-compatibility,
 \begin{equation}\label{Chern connection}
 \Gamma^{\alpha}_{\mu\nu}=\gamma^{\alpha}_{\mu\nu}-g^{\alpha\lambda}\left(A_{\lambda\mu\beta}\frac{N^\beta_\nu}{F}-A_{\mu\nu\beta}\frac{N^\beta_\lambda}{F}+A_{\nu\lambda\beta}\frac{N^\beta_\mu}{F}\right),
 \end{equation}
 where $\gamma^{\alpha}_{\mu\nu}$ is the formal Christoffel symbols of the
second kind with the same form of Riemannian connection, $N^\mu_\nu$
is defined as
$N^\mu_\nu\equiv\gamma^\mu_{\nu\alpha}y^\alpha-A^\mu_{\nu\lambda}\gamma^\lambda_{\alpha\beta}y^\alpha
y^\beta$
 and $A_{\lambda\mu\nu}\equiv\frac{F}{4}\frac{\partial}{\partial y^\lambda}\frac{\partial}{\partial y^\mu}\frac{\partial}{\partial y^\nu}(F^2)$ is the
 Cartan tensor (regarded as a measurement of deviation from the Riemannian
 Manifold). In terms of Chern connection, the curvature of Finsler space is given as
\begin{equation}\label{Finsler curvature}
R^{~\lambda}_{\kappa~\mu\nu}=\frac{\delta
\Gamma^\lambda_{\kappa\nu}}{\delta x^\mu}-\frac{\delta
\Gamma^\lambda_{\kappa\mu}}{\delta
x^\nu}+\Gamma^\lambda_{\alpha\mu}\Gamma^\alpha_{\kappa\nu}-\Gamma^\lambda_{\alpha\nu}\Gamma^\alpha_{\kappa\mu},
\end{equation}
where $\frac{\delta}{\delta x^\mu}=\frac{\partial}{\partial x^\mu}-N^\nu_\mu\frac{\partial}{\partial y^\nu}$.

The gravity in Finsler spacetime has been investigated for a long
time \cite{Takano,Ikeda,Tavakol1, Bogoslovsky1}. In this paper, we introduce vacuum field equation by the way discussed first by Pirani \cite{Pirani, Rutz}. In Newton's theory of gravity, the equation of motion of a test particle is given as
\begin{equation}
\label{dynamic Newton}
\frac{d^2x^i}{dt^2}=-\eta^{ij}\frac{\partial \phi}{\partial x^i},
\end{equation}
where $\phi=\phi(x)$ is the gravitational potential and $\eta^{ij}$ is Euclidean metric. For an infinitesimal transformation $x^i\rightarrow x^i+\epsilon\xi^i$($|\epsilon|\ll1$), the equation (\ref{dynamic Newton}) becomes, up to first order in $\epsilon$,
\begin{equation}
\label{dynamic Newton1}
\frac{d^2x^i}{dt^2}+\epsilon\frac{d^2\xi^i}{dt^2}=-\eta^{ij}\frac{\partial \phi}{\partial x^i}-\epsilon\eta^{ij}\xi^k\frac{\partial^2\phi}{\partial x^j\partial x^k}.
\end{equation}
Combining the above equations(\ref{dynamic Newton}) and (\ref{dynamic Newton1}), we obtain
\begin{equation}
\frac{d^2\xi^i}{dt^2}=\eta^{ij}\xi^k\frac{\partial^2\phi}{\partial x^j\partial x^k}\equiv\xi^kH^i_k.
\end{equation}
In Newton's theory of gravity, the vacuum field equation is given as $H^i_i=\bigtriangledown^2\phi=0$. It means that the tensor $H^i_k$ is traceless in Newton's vacuum.

In general relativity, the geodesic deviation gives similar equation
\begin{equation}
\frac{D^2\xi^\mu}{D\tau^2}=\xi^\nu \tilde{R}^\mu_{~\nu},
\end{equation}
where $\tilde{R}^\mu_{~\nu}=\tilde{R}^{~\mu}_{\lambda~\nu\rho}\frac{dx^\lambda}{d\tau}\frac{dx^\rho}{d\tau}$. Here, $\tilde{R}^{~\mu}_{\lambda~\nu\rho}$ is Riemannian curvature tensor, $D$ denotes the covariant derivative alone the curve $x^\mu(\tau)$. The vacuum field equation in general relativity gives $\tilde{R}^{~\lambda}_{\mu~\lambda\nu}=0$\cite{Weinberg}. It implies that the tensor $\tilde{R}^\mu_{~\nu}$ is also traceless, $\tilde{R}\equiv\tilde{R}^\mu_{~\mu}=0$.

In Finsler spacetime, the geodesic deviation gives \cite{Book by Bao}
\begin{equation}
\frac{D^2\xi^\mu}{D\tau^2}=\xi^\nu R^\mu_{~\nu},
\end{equation}
where $R^\mu_{~\nu}=R^{~\mu}_{\lambda~\nu\rho}\frac{dx^\lambda}{d\tau}\frac{dx^\rho}{d\tau}$. Here, $R^{~\mu}_{\lambda~\nu\rho}$ is Finsler curvature tensor defined in (\ref{Finsler curvature}), $D$ denotes covariant derivative $\frac{D\xi^\mu}{D\tau}=\frac{d\xi^\mu}{d\tau}+\xi^\nu\frac{dx^\lambda}{d\tau}\Gamma^\mu_{\nu\lambda}(x,\frac{dx}{d\tau})$. Since the vacuum field equations of Newton's gravity and general relativity have similar form, we may assume that vacuum field equation in Finsler spacetime hold similar requirement as the case of Netwon's gravity and general relativity. It implies that the tensor $R^\mu_{~\nu}$ in Finsler geodesic deviation equation should be traceless, $R\equiv R^\mu_{~\mu}=0$. In fact, we have proved that the analogy from the geodesic deviation equation is valid at least in Finsler spacetime of Berwald type \cite{Finsler DM}. We suppose that this analogy is still valid in general Finsler spacetime.

It should be noticed that $H$ is called the Ricci scaler, which is a geometrical invariant. For a tangent plane $\Pi\subset T_xM$ and a non-zero vector $y\in T_xM$, the flag curvature is defined as
\begin{equation}
\label{flag curvature}
K(\Pi,y)\equiv\frac{g_{\lambda\mu}R^\mu_{~\nu}u^\nu u^\lambda}{F^2g_{\rho\theta}u^\rho u^\theta-(g_{\sigma\kappa}y^\sigma u^\kappa)^2},
\end{equation}
where $u\in\Pi$. The flag curvature is a geometrical invariant and a generalization of the sectional curvature in Riemannian geometry. It is clear that the Ricci scaler $R$ is the trace of $R^\mu_{~\nu}$, which is the predecessor of flag curvature. Therefore, the value of Ricci scaler $R$ is invariant under the coordinate transformation. Furthermore, the predecessor of flag curvature could be written in terms of the geodesic spray coefficient
\begin{equation}\label{predecessor flag curvature}
R^\mu_{~\nu}=2\frac{\partial G^\mu}{\partial x^\nu}-y^\lambda\frac{\partial^2 G^\mu}{\partial x^\lambda\partial y^\nu}+2G^\lambda\frac{\partial^2 G^\mu}{\partial y^\lambda\partial y^\nu}-\frac{\partial G^\mu}{\partial y^\lambda}\frac{\partial G^\lambda}{\partial y^\nu}.
\end{equation}
Thus, the Ricci scaler $R$ is insensitive to specific connection form. It only depends on the length element $F$. The gravitational vacuum field equation $R=0$ is universal in any types of theories of Finsler gravity. Pfeifer {\it et al}. \cite{Pfeifer1} have constructed gravitational dynamics for Finsler spacetimes in terms of an action integral on the unit tangent bundle. Their results also show that the gravitational vacuum field equation in Finsler spacetime is $R=0$.

\section{The Newtonian limit in Finsler spacetime}
It is well known that the Minkowski spacetime is a trivial solution of Einstein's vacuum field equation. In the Finsler spacetime, the trivial solution of Finslerian vacuum field equation is called locally Minkowski spacetime. A Finsler spacetime is called a locally Minkowshi spacetime if there is a local coordinate system $(x^\mu)$, with induced tangent space coordinates $y^\mu$, such that $F$ depends only on $y$ and not on $x$. Using the formula (\ref{predecessor flag curvature}), one knows obvious that locally Minkowski spacetime is a solution of Finslerian vacuum field equation.

In Ref. \cite{Finsler GW}, we supposed that the metric is close to the locally Minkowski metric $\eta_{\mu\nu}(y)$,
\begin{equation}
g_{\mu\nu}=\eta_{\mu\nu}(y)+h_{\mu\nu}(x,y),~~|h_{\mu\nu}|\ll1
\end{equation}
and found that the gravitational vacuum field equation is of the form
\begin{equation}\label{field equation}
\eta^{\mu\nu}\frac{\partial^2 h_{\alpha\beta}}{\partial x^\mu \partial x^\nu}
-\frac{1}{2}\frac{\partial \eta^{\mu\nu}}{\partial y^\mu}\frac{\partial^2 h_{\alpha\beta}}{\partial x^\nu \partial x^\lambda}y^\lambda =0,
\end{equation}
where the lowering and raising of indices are carried out by $\eta_{\mu\nu}$ and its matrix inverse $\eta^{\mu\nu}$.
To make contact with Newton's gravity, we consider the gravitational field $h_{\mu\nu}$ is stationary and particle moving very slowly. In general relativity, gravitational field $h_{00}$ corresponds to the Newton's gravitational potential. Here, we consider that only $h_{00}$ is the non vanish components of $h_{\mu\nu}$. Then, we obtain from (\ref{field equation}) that
\begin{equation}\label{static field eq}
\eta^{ij}\frac{\partial^2 h_{00}}{\partial x^i \partial x^j}=0.
\end{equation}
The solution of (\ref{static field eq}) is given as
\begin{equation}
h_{00}=\frac{C}{R}\eta_{00},~~ (R\neq0),
\end{equation}
where $C$ is a constant and $R^2=\eta_{ij}x^i x^j$. The gravitational theory should reduce to general relativity, if the Finsler metric $g_{\mu\nu}$ reduces to Riemann metric. Thus, for a giving gravitational source with mass $M$ at $R=0$, the gravitational field equation should be of the form
\begin{equation}\label{static field eq1}
\eta^{ij}\frac{\partial^2 h_{00}}{\partial x^i \partial x^j}=8\pi_F G\rho\eta_{00},
\end{equation}
where $\rho$ is the energy density of gravitational source. In Finsler spacetime, the space volume of $\eta_{\mu\nu}(y)$ \cite{Book by Bao} is different the one in Euclidean space. We used $\pi_F$ in (\ref{static field eq1}) to represent the difference, where 
\begin{equation}
\pi_F\equiv\frac{3}{4}\int_{R=1}\sqrt{g}dx^1\wedge dx^2\wedge dx^3
\end{equation} 
and $g$ is the determinant of $\eta_{ij}$.
The solution of (\ref{static field eq1}) is given as
\begin{equation}\label{gravitational poten}
h^0_0=\frac{2GM}{R}.
\end{equation}
Here, we have used $h^0_0$ to denote the gravitational field instead of $h_{00}$. It is due to the fact that $\eta_{00}\neq1$ and we want to obtain the formula (\ref{gravitational poten}) which insensitive to the spacetime index.

In the approximation of Newton's limit, the geodesic equation (\ref{geodesic}) reduces to
\begin{eqnarray}\label{geodesic compo1}
\frac{d^2x^0}{d\tau^2}-\frac{\eta^{0i}}{2}\frac{\partial h_{00}}{\partial x^i}\frac{dx^0}{d\tau}\frac{dx^0}{d\tau}&=&0,\\
\label{geodesic compo2}
\frac{d^2x^i}{d\tau^2}-\frac{\eta^{ij}}{2}\frac{\partial h_{00}}{\partial x^j}\frac{dx^0}{d\tau}\frac{dx^0}{d\tau}&=&0.
\end{eqnarray}
The equation (\ref{geodesic compo1}) implies that $\frac{dx^0}{d\tau}$ is a function of $h_{00}$. Since $|h_{00}|\ll1$, $\frac{dx^0}{d\tau}$ could be treated as a constant in equation (\ref{geodesic compo2}). Then, we find from (\ref{geodesic compo2}) that
\begin{equation}\label{law of gravity}
\frac{d^2x^i}{{dx^0}^2}=-\frac{GM}{R^2}\frac{x^i}{R},
\end{equation}
where ${dx^0}^2=\eta_{00}dx^0 dx^0$. The formula (\ref{law of gravity}) means that the law of gravity in Finsler spacetime is similar to the Newton's gravity. The difference is that space length is Finslerian. It is what we expect from Finslerian gravity, because the length difference is the major attribute of Finsler geometry.

The Newton's gravity or general relativity is compatible with the astronomical tests in solar system. It means that the Minkowski spacetime well describes the physics in solar system. It may be expected that the Finsler gravity (\ref{law of gravity}) domains in large scale. The empirical Tully-Fisher relation \cite{TF} between galaxy luminosity and rotational speed implies that the rotational speed of galaxy has a limitation. It hints that we may consider the space length of galaxies to be Finslerian
\begin{equation}\label{length gala1}
\eta_{ij}=\delta_{ij}\sqrt{1-\left(\frac{GMa_0{y^0}^4}{(\delta_{mn}y^m y^n)^2}\right)^2},
\end{equation}
where $a_0$ is the constant of MOND. In Finsler spacetime, the speed of particle is given as  $v^i\equiv\frac{dx^i}{dx^0}=\frac{y^i}{y^0}$. The space length of galaxies (\ref{length gala1}) could be write into a concise form
\begin{equation}\label{length gala2}
R=r\sqrt{1-\left(\frac{GMa_0}{v^4}\right)^2},
\end{equation}
where $r^2\equiv\delta_{ij}x^i x^j$ and $v^2\equiv\delta_{ij}v^i v^j$. By making use of formula (\ref{law of gravity}), we obtain the approximate dynamical equation in galaxy system. It is given as
\begin{equation}\label{law of gravity1}
\frac{GM}{R^2}=\frac{v^2}{R}.
\end{equation}
Substituting formula (\ref{length gala2}) into equation (\ref{law of gravity1}), we obtain that
\begin{equation}\label{law of gravity2}
\frac{GM}{r\sqrt{1-\left(GMa_0/v^4\right)^2}}=v^2.
\end{equation}
One could find from (\ref{law of gravity2}) that the rotational speed of galaxies has an asymptotic value
\begin{equation}
\lim_{r\rightarrow\infty}v^4=GMa_0.
\end{equation}
The asymptotic speed stems from the Tully-Fisher relation. The formula (\ref{law of gravity2}) could be written into a familiar form
\begin{equation}\label{MOND}
\frac{GM}{r^2}=\frac{v^2}{r}\mu\left(\frac{v^2}{ra_0}\right),
\end{equation}
where $\mu(x)=x/\sqrt{x^2+1}$ is interpolating function of MOND. It indicates that the law of gravity (\ref{law of gravity2}) in Finsler spacetime is MOND and the spacetime structure of MOND is ``Tully-Fisher" like (\ref{length gala2}). It should be noticed that the formula (\ref{law of gravity2}) is given in natural units ($c=1$). The term $L_g\equiv GM/c^2$ corresponds to the typical galaxy scale where the gravity of galaxy is dominated. The term $L_0\equiv c^2/a_0\approx2\pi L_H\approx10^{29} {\rm cm}$\cite{MOND}, where $L_H$ is the Hubble radius. Thus, the constant term $GMa_0$ in (\ref{law of gravity2}) equals ratio of galaxy scale $L_g$ to cosmological scale $L_0$. The law of gravity (\ref{law of gravity2}) in Finsler spacetime hints that there is connection between MOND and cosmology.

\section{Conclusions}
In this paper, we presented a Finsler geometry origin of MOND. We showed that the spacetime structure of galaxies may be Finslerian (\ref{length gala2}). The ``Tully-Fisher" like length (\ref{length gala2}) could be the reason of the empirical Tully-Fisher relation. The law of gravity in galaxies was shown as formula (\ref{law of gravity2}). It hints that there is connection between MOND and cosmology.

The strong and weak gravitational lensing observations of Bullet Cluster 1E0657-558\cite{Bullet} could not be explained well by MOND and Bekenstein's relativistic version of MOND\cite{Angus}. The surface density $\Sigma$-map and the convergence $\kappa$-map of Bullet Cluster 1E0657-558 show that the center of baryonic matters separates from the center of gravitational force, and the distribution of gravitational force do not possess spherical symmetry. One should notice that Finsler metric (\ref{length gala1}) satisfies
\begin{equation}\label{abso homo}
\eta_{ij}(-y)=\eta_{ij}(y).
\end{equation}
Most of galaies possess spherical symmetry, so length element should satisfy (\ref{abso homo}). There is a class of Finsler spacetime that do not satisfy (\ref{abso homo}). For example, the Randers spacetime \cite{Randers}
\begin{equation}
F(x,y)=\sqrt{a_{\mu\nu}y^\mu y^\nu}+b_\mu y^\mu~.
\end{equation}
The space structure of Bullet Cluster may be one of this class, and its Finslerian gravitational behavior may account for the observations of Bullet Cluster.

\vspace{1cm}
\begin{acknowledgments}
We would like to thank M. H. Li, S. Wang, Y. G. Jiang and H. N. Lin for useful discussions. The work was supported by the NSF of China under Grant No. 11075166 and 11147176.
\end{acknowledgments}

\end{document}